Perfect nonradiating modes in dielectric nanofiber with elliptical cross-section

V. V. Klimov[1*] and D. V. Guzatov[2]

[1]P. N. Lebedev Physical Institute, Russian Academy of Sciences, 53 Leninsky Prospekt, Moscow 119991, Russia,

[2]Yanka Kupala State University of Grodno, 22 Ozheshko str., Grodno 230023, Belarus

**Abstract**

The existence of an infinite number of perfect nonradiating modes in elliptical nanofibers is demonstrated. Dispersion laws are found for TM and TE perfect modes in circular and elliptical waveguides with an arbitrary eccentricity. Numerical simulations in Comsol Multiphysics have shown that these modes can be excited by plane waves of a certain parity and have extremely low radiative losses and extremely high quality factors. The found modes can be used to create highly sensitive nanosensors and other optical nanodevices where radiation losses should be minimal.

## 1. INTRODUCTION

At present, the properties of dielectric nanoparticles with a high refractive index and low radiative losses are being investigated actively. The physics of optical phenomena in such nanoparticles is very complicated and leads to many interesting applications, such as nano-antennas [1-3], nanolasers [4,5] invisible Mie scatterers [6], and nonlinear nanophotonics [7]. As in any other field of physics, all these phenomena are associated with the existence of certain eigenmodes in nanoparticles.

[*] klimov256@gmail.com

For applications, modes with strong field localization and low radiation losses are of particular interest. This kind of modes have attracted the closest attention of the leading scientific groups, discovered several types of weakly radiating phenomena: bound states in the continuum [8-13], anapole states [7,8, 14-20], supercavity modes [5, 21,22], pseudo-modes [23].

Usually, the eigenmodes are found by solving sourceless Maxwell's equations with the Sommerfeld radiation conditions at infinity [24], and therefore such modes are fundamentally related to radiation losses. Moreover, such modes grow unlimitedly at infinity, requiring the development of very complex artificial approaches for their use, see e.g. [25-28].

However, finding all the modes is a non-trivial task, not only from a computational point of view, and the modes investigated in the abovementioned works do not exhaust the entire set of modes that exist in dielectric particles.

It was shown in works [29,30] that an infinite number of perfect nonradiating eigenmodes can exist in 3D dielectric nanoparticles of a finite volume. More specifically, in [29,30] to find new eigenmodes it was suggested to look for the electromagnetic fields outside the particle in the form of a superposition of solutions of the Maxwell's equations that are nonsingular in unlimited free space, including the interior of the nanoparticle. This approach is fundamentally different from the usual one assuming that the functions describing the fields outside the body can have singularities upon analytic continuation into its interior. For example, expanding any field component $E^{out}(r,\theta,\varphi,\omega)$ (where $r,\theta,\varphi$ are spherical coordinates, $\omega$ is the frequency) outside the resonator over spherical harmonics $Y_n^m(\theta,\varphi)$ in accordance with the Sommerfeld radiation condition, usually one uses spherical Hankel functions $h_n^{(1)}(k_0 r)$ (here $k_0 = \omega/c$ and $c$ is the velocity of light) singular at $r = 0$ (inside the resonator) and fundamentally related to radiation [24]:

$$E^{out}(r,\theta,\varphi,\omega) = \sum_{n,m} A_{nm} h_n^{(1)}(k_0 r) Y_n^m(\theta,\varphi),$$
(1)

where $A_{nm}$ are the expansion coefficients. Outside the nanoparticle, however, it was proposed to use the solutions of Maxwell equations *nonsingular at the origin* to find the perfect nonradiating modes. For example, one can look for fields outside the resonator in the form:

$$E(r,\theta,\varphi,\omega) = \sum_{n,m} A_{nm} j_n(k_0 r) Y_n^m(\theta,\varphi), \quad (2)$$

where $j_n(k_0 r)$ are nonsingular spherical Bessel functions. Definitions and properties of the special functions mentioned in the present work can be found, for example, in [31].

Obviously, if a solution of the Maxwell's equations with outside fields in the form (2) exists, then, in principle, it does not have a flux of energy and radiation. Since (2) has no singularities in the entire space (including the nanoparticle interior), finding of modes in the infinite space can be reduced to finding fields in the volume of the nanoparticle only. As a result, the system of equations that determine unambiguously the perfect nonradiating modes in a nonmagnetic nanoparticle with the permittivity $\varepsilon$ can be written as a system of two equations for two independent auxiliary fields $\mathbf{E}_1$, $\mathbf{E}_2$:

$$\begin{aligned}
\nabla \times \mathbf{E}_1 &= ik_0 \mathbf{H}_1, & \nabla \times \mathbf{H}_1 &= -ik_0 \varepsilon \mathbf{E}_1, & \text{inside nanoparticle} \\
\nabla \times \mathbf{E}_2 &= ik_0 \mathbf{H}_2, & \nabla \times \mathbf{H}_2 &= -ik_0 \mathbf{E}_2, & \text{inside nanoparticle}, \\
\mathbf{n} \times (\mathbf{E}_1 - \mathbf{E}_2) &= 0, & \mathbf{n} \times (\mathbf{H}_1 - \mathbf{H}_2) &= 0, & \text{at the boundary}
\end{aligned} \quad (3)$$

where $\nabla$ is the gradient operator, $\mathbf{n}$ is the unit normal vector to the boundary. It is important to note that the system (3) is self-sufficient, and there is no need to impose any condition at infinity to solve it.

At some real values of frequency or wavenumber $k_0$, the system of equations (3) becomes compatible, that is, the perfect nonradiating modes appear. It is very important that due to a specific structure of (3) there is nothing common between the frequencies of perfect modes and the frequencies of usual quasinormal modes. The modes found from (3) are orthogonal in the sense that

$$\int_V dV \left( \varepsilon \mathbf{E}_{1,n} \cdot \mathbf{E}_{1,m} - \mathbf{E}_{2,n} \cdot \mathbf{E}_{2,m} \right) = \delta_{nm}, \quad (4)$$

where the integration is over the nanoparticle volume $V$, and $\delta_{nm}$ is the Kronecker's delta. The condition (4) is also drastically different from orthogonality conditions for usual normal modes.

The physical fields inside the nanoparticle are determined by $\mathbf{E}_1$, $\mathbf{H}_1$, while the physical fields outside the particle are determined by the analytic continuation of the $\mathbf{E}_2$, $\mathbf{H}_2$.

The system of equations (4) is very complicated, and a rigorous mathematical theory does not yet exist for it in a general case. Nevertheless, in [29,30] the conditions for the existence of perfect nonradiating modes for arbitrary spheres, spheroids, and superspheroids, describing well practically all forms of nanoparticles interesting for applications, were found.

In this paper, we generalize the approach of [29,30] to the case of dielectric nanofibers of elliptical cross-section with zero propagation constant, that is, actually we will consider a 2D problem. In this case, in contrast to the formulation of problems for finding bound states in the continuum (BIC [32-34]), assuming one or another periodic structure of waveguides, we will consider straight cylindrical waveguides of a circular or elliptical cross-section. In this case, in accordance with the approach proposed in [29,30] we will use functions that are nonsingular in the entire space, that is, inside the waveguide, as functions describing the fields outside the waveguide. The problem geometry is shown in FIG. 1.

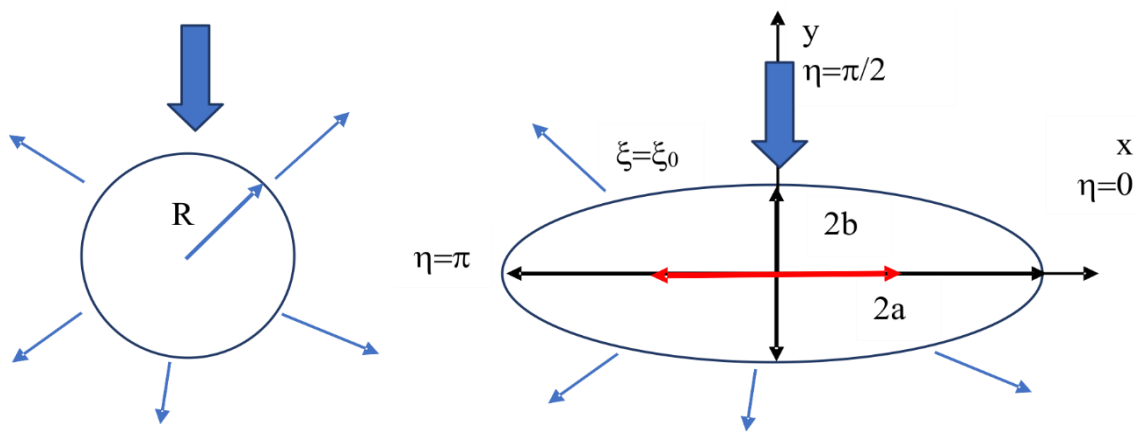

FIG. 1. (Color online) Geometry of the problem: (a) a circular waveguide cross-section, (b) a waveguide with an elliptical cross-section. The blue arrow designates an excitation field.

The plan of the rest part of the work is as follows. Section 2 will present analytical solutions for perfect nonradiating modes in a circular waveguide and numerical experiments in Comsol Multiphysics confirming the presence of deep minima in the scattered power at perfect nonradiating modes frequencies, that can be verified experimentally. Section 3 contains analytical solutions for perfect nonradiating TM modes in a waveguide of an arbitrary elliptical cross-section and the results of numerical experiments in Comsol Multiphysics, confirming their practical importance. In Section 4 the analytical solutions for perfect nonradiating TE modes in a waveguide of an arbitrary elliptical cross-section will be considered. Section 5 will give an example of applying of the found effects to design of displacement sensors.

## 2. PERFECT NONRADIATING MODES IN A CYLINDRICAL WAVEGUIDE WITH A CIRCULAR CROSS-SECTION

The existence of perfect nonradiating modes in a cylindrical waveguide is easiest to prove for a circular cross-section (FIG. 1a).

### 2.1 TM polarization

In the case of TM polarization, the magnetic field has only one z-component $H_z$, in the direction along the axis of the waveguide. For $H_z$ one can write the Helmholtz equation [35]:

$$\left(\frac{\partial^2}{\partial x^2} + \frac{\partial^2}{\partial y^2} + k_0^2 \varepsilon(x,y)\right) H_z = 0, \qquad (5)$$

In the case under consideration, it is convenient to solve eq. (5) in the cylindrical coordinates $0 \leq \rho < \infty$, $0 \leq \varphi \leq 2\pi$. Inside the waveguide of the radius $R$, an arbitrary *nonsingular* solution has the form:

$$H_z^{in} = J_m\left(k_0 \rho \sqrt{\varepsilon}\right) e^{im\varphi}, \quad \rho < R, \qquad (6)$$

while outside the waveguide, an arbitrary nonsingular solution in the entire space has the form:

$$H_z^{out} = J_m\left(k_0 R\sqrt{\varepsilon}\right)\frac{J_m(k_0\rho)}{J_m(k_0 R)}e^{im\varphi}, \quad \rho > R. \tag{7}$$

By construction, (6) and (7) are continuous on the surface of the waveguide. The requirement for the continuity of the tangential $\varphi$-component of the electric field strength $E_\varphi$ leads to the dispersion equation

$$J_m\left(k_0 R\sqrt{\varepsilon}\right) = \frac{1}{\sqrt{\varepsilon}} J_m(k_0 R)\frac{J'_m\left(k_0 R\sqrt{\varepsilon}\right)}{J'_m(k_0 R)}, \tag{8}$$

where the prime next to the function means its derivative. This equation is different from the equation that defines usual modes [35]:

$$J_m\left(k_0 R\sqrt{\varepsilon}\right) = \frac{1}{\sqrt{\varepsilon}} H_m^{(1)}(k_0 R)\frac{J'_m\left(k_0 R\sqrt{\varepsilon}\right)}{H_m^{(1)\prime}(k_0 R)} \tag{9}$$

by the Hankel functions replacement with the usual Bessel functions.

It is easy to see that, in the limit $k_0 R \to \infty$, eq. (8) reduces to the equation

$$\begin{gathered}(-1)^m N^- \cos\left(N^+ k_0 R\right) + N^+ \sin\left(N^- k_0 R\right) = 0, \\ N^\pm = \sqrt{\varepsilon} \pm 1 \end{gathered} \tag{10}$$

whence it follows that equation (8) as the function of $k_0 R$ definitely has an infinite number of real roots determining the frequencies of perfect nonradiating modes. In this case, the solution (6), (7) is a real function.

Figure 2 shows the dependences of $H_z$ on the radius for an usual and perfect nonradiating mode (6), (7).

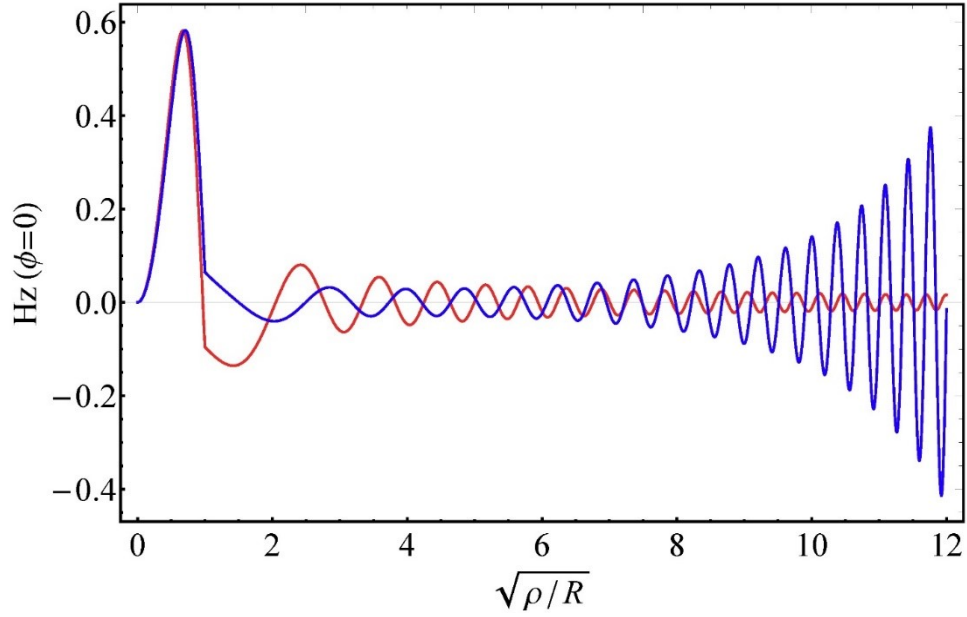

FIG. 2. (Color online) Dependence of Re($H_z(\varphi=0)$) on the radius for the perfect nonradiating mode ((6), (7), red color, $k_0R = 0.912036$) and for the usual mode (blue color, $k_0R = 0.821076 - i0.0297689$) in a cylinder with $\varepsilon=20$.

From FIG. 2 it is clearly seen that the perfect nonradiating mode decreases at infinity. This fact radically distinguishes it from the usual (quasinormal) mode, which increases without limit at infinity.

FIG. 3 shows the dependence of the frequencies of TM perfect nonradiative modes (PTM) $k_0R\sqrt{\varepsilon}$ in a cylindrical waveguide of circular cross-section on $1/\varepsilon$.

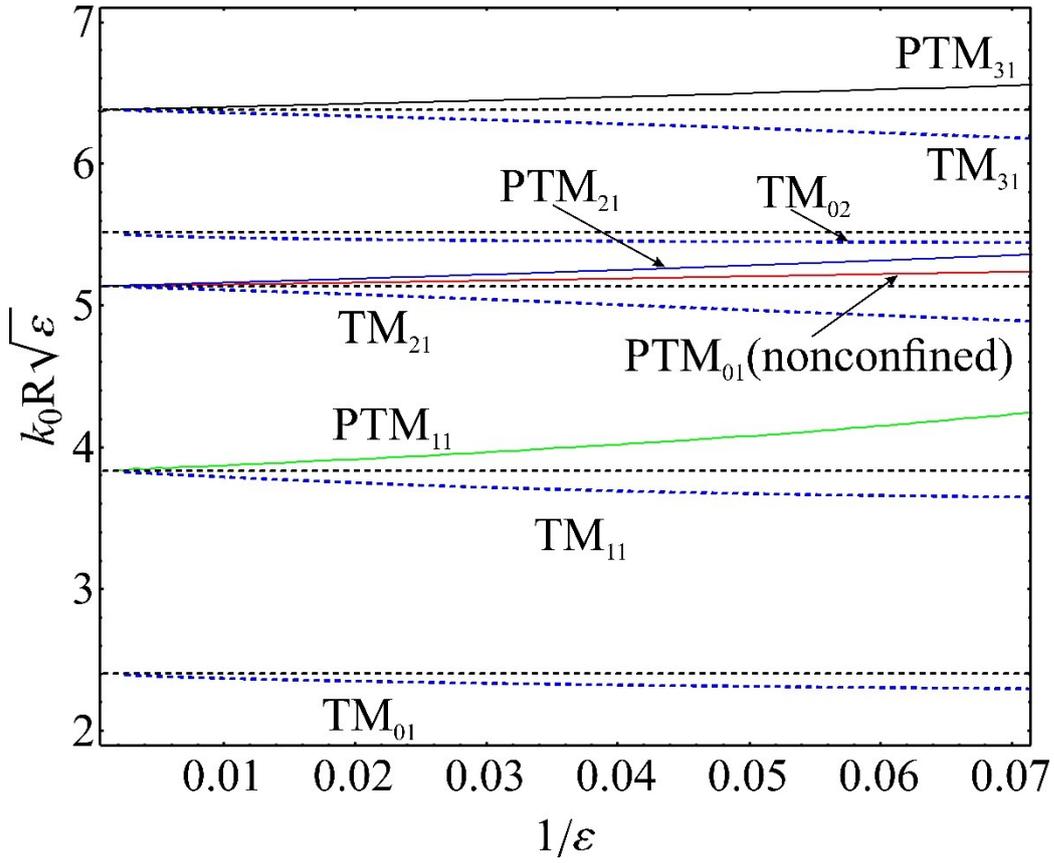

FIG. 3. (Color online) The dependence of frequencies of the TM perfect nonradiating modes (PTM) on $1/\varepsilon$. The black horizontal dashed lines show the roots of the Bessel functions. The blue dashed lines show the position of the usual quasinormal TM modes. The first index $m$ at the bottom of the TM$_{mn}$ or PTM$_{mn}$ indicates the order of the mode, and the second index $n$ indicates n$^{th}$ root of the (8) for PTM modes and (9) for TM modes.

It follows from FIG. 3, that the frequency of the PTM$_{01}$ mode in the limit $\varepsilon \to \infty$ is determined by the root of the Bessel function $J_2\left(k_0 R\sqrt{\varepsilon}\right) = 0$, and not by the root of $J_0\left(k_0 R\sqrt{\varepsilon}\right) = 0$. It indicates that this mode, unlike the others, is not confined [36], that is, its field outside the waveguide in this limit does not disappear.

Indeed, the solution of the Maxwell equations for this mode at $\varepsilon \to \infty$ can be written as

$$H_z = \begin{cases} J_0\left(j_{2,1}\dfrac{\rho}{R}\right), & \rho < R \\ J_0\left(j_{2,1}\dfrac{\rho}{R\sqrt{\varepsilon}}\right), & \rho > R \end{cases}, \quad (11)$$

where $j_{2,1}$ is the first root of the equation $J_2(z) = 0$. It is immediately clear from (11) that the PTM$_{01}$ mode is not confined and does not decrease at infinity (but does not grow like usual modes). This situation is specific for 2D geometry, where the dispersion equation (8) at $m=0$ in the limit $\varepsilon \to \infty$ takes the form:

$$J_2\left(k_0 R\sqrt{\varepsilon}\right) = 0. \quad (12)$$

For all other perfect modes, the dispersion equation (8) in the case $\varepsilon \to \infty$ takes the form:

$$J_m\left(k_0 R\sqrt{\varepsilon}\right) = 0, \quad m \neq 0 \quad (13)$$

providing the vanishing of the field outside the waveguide (7), i.e., in this limit, all perfect nonradiating TM modes with $m \neq 0$ are confined.

Note that for perfect modes in 3D axisymmetric particles, all TM and PTM modes are confined.

The found perfect nonradiating modes are not abstract solutions of sourceless Maxwell's equations. They are of great practical importance for finding the conditions for extremely small or even zero scattered power at a finite stored energy, leading to the unlimited radiative $Q$ factor.

It is easy to see that the excitation of a perfect nonradiating mode (6)-(8) by a cylindrical wave proportional to (7) leads to zero scattered power, since the condition (8) corresponds to zero of the Mie scattering coefficient [24].

Indeed, if the excitation field is taken in the form:

$$H_z^{exc} = J_m(k_0 \rho) e^{im\varphi}, \quad (14)$$

then the scattered field outside the waveguide will have the form:

$$H_z^{scat} = a_m H_m^{(1)}(k_0\rho)e^{im\varphi},$$

$$a_m = -\frac{\sqrt{\varepsilon}J_m(k_0R\sqrt{\varepsilon})J'_m(k_0R) - J_m(k_0R)J'_m(k_0R\sqrt{\varepsilon})}{\sqrt{\varepsilon}J_m(k_0R\sqrt{\varepsilon})H_m^{(1)'}(k_0R) - H_m^{(1)}(k_0R)J'_m(k_0R\sqrt{\varepsilon})}, \quad (15)$$

where $a_m$ is the TM Mie scattering coefficients for the circular cylinder (see [24], p. 199).

When the condition (8) is satisfied, the scattered field (15) and the scattered power are equal to zero!

The realization of a cylindrical standing wave (14) in an experiment is a difficult task, and therefore the scattering of simple standing plane waves by a circular cylinder have been simulated numerically within Comsol Multiphysics software. Results of the simulation are shown in FIG. 4. These results are in excellent agreement with the analytical solution [37], which confirms the reliability of using Comsol to find perfect nonradiating modes within more complicated geometries (see Sections 3,4).

FIG.4 shows clearly the appearance of deep minima in the spectra of the scattered power and extremely high maxima in the spectra of the radiative $Q$ factor, respectively. These extremums demonstrate clearly the appearance of high-quality perfect $PTM_{11}$ and $PTM_{01}$ modes. Note that $Q$ factor of the $PTM_{11}$ mode is by six orders of magnitude greater than $Q$ factor of a usual $TM_{11}$ mode!

Figure 5 shows the spatial distribution of the excitation field

$$H_z^{exc} = \sin(k_0 x \cos\alpha)\cos(k_0 y \sin\alpha) \quad (16)$$

without nanofiber (left) and the full field in the presence of nanofiber (right) at the frequency $f=2.642\times10^{14}$ Hz of the perfect nonradiating $PTM_{11}$ mode (see FIG. 4).

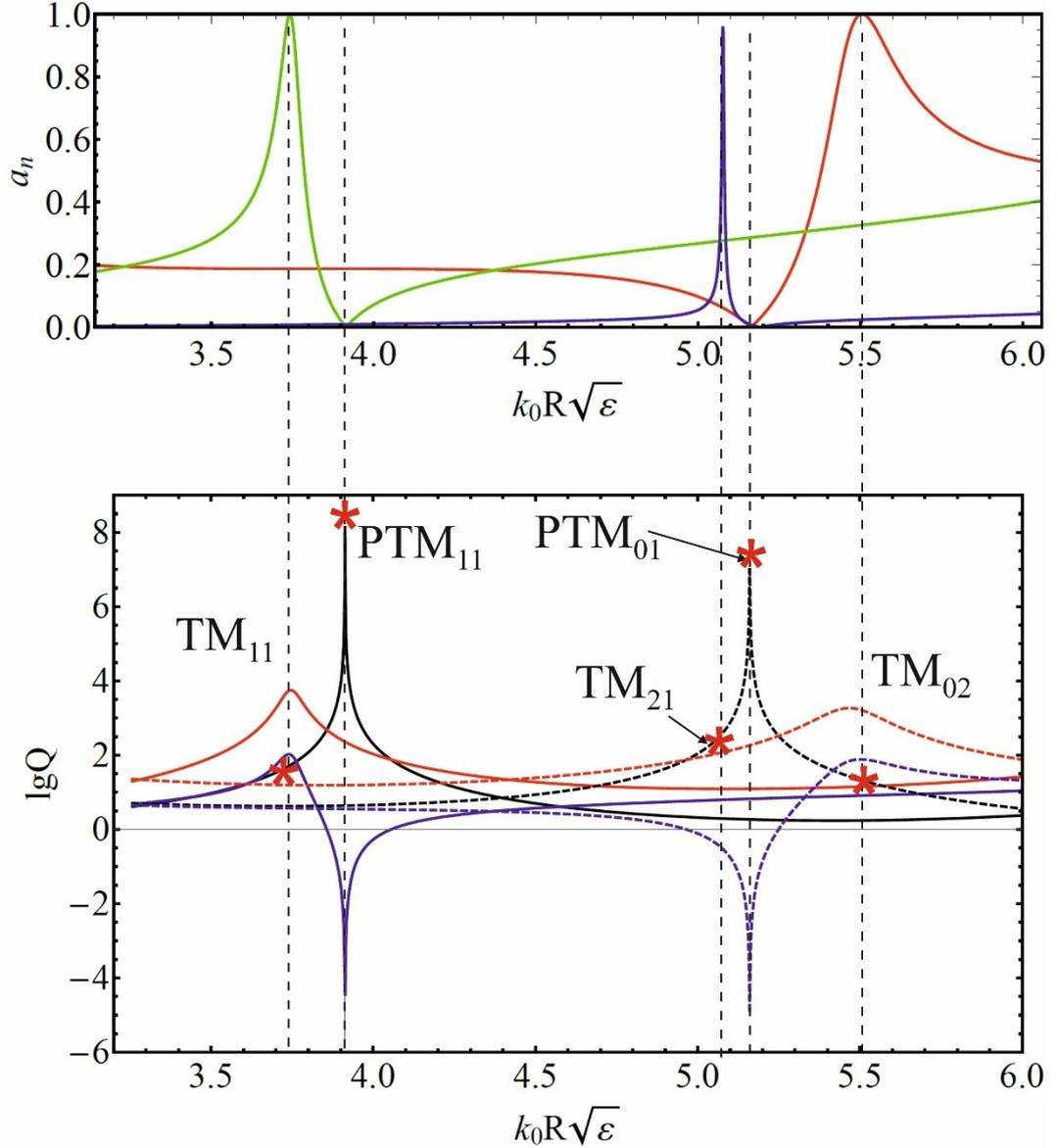

FIG. 4. (Color online) The dependence of (a) TM Mie coefficients $a_n$ and (b) scattered power $\lg P^{scat}$ (blue), stored energy $\lg W^{stored}$ (red), and radiative $Q$ factor $\lg Q$ (black) on the size parameter of a waveguide of the circular cross-section ($\varepsilon=50$) obtained within the Comsol simulation. The solid curves correspond to the excitation field $H_z^{exc} = \sin(k_0 x \cos\alpha)\cos(k_0 y \sin\alpha)$, and the dashed curves correspond to the excitation field $H_z^{exc} = \cos(k_0 x \cos\alpha)\cos(k_0 y \sin\alpha)$, where $\alpha=\pi/4$. The asterisks on the black curves show the $Q$ factor values of usual and perfect modes. PTM stands for perfect TM nonradiating mode.

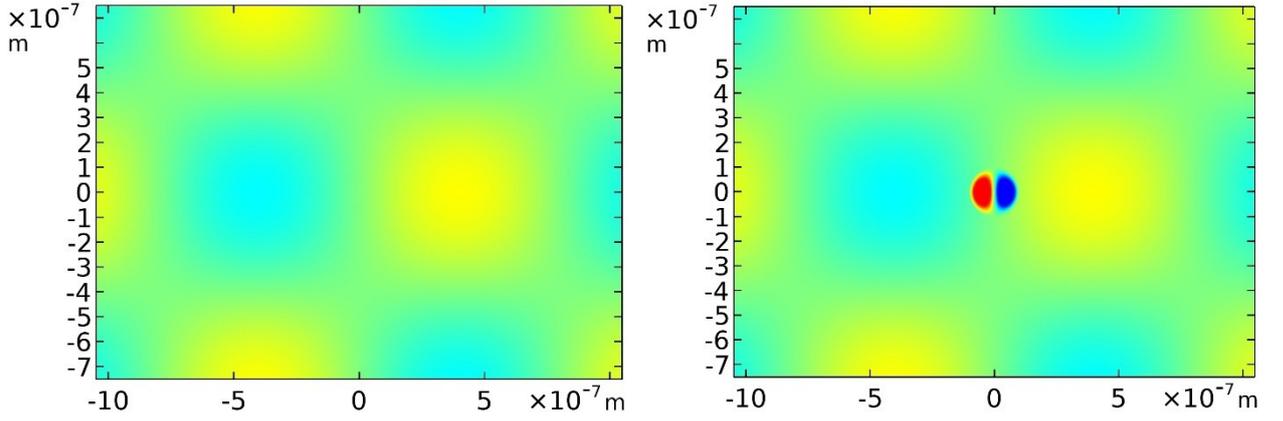

FIG. 5. (Color online) Spatial distribution of the excitation field (16) without nanofiber (left) and full field in the presence of nanofiber (right) at the frequency $f=2.642\times10^{14}$ Hz of the perfect nonradiating mode $PTM_{11}$ at $R=100$ nm, $\varepsilon=50$, and $\alpha=\pi/4$.

Figure 5 shows clearly that the mounting of a nanowaveguide into the excitation field (16) does not affect the field outside it, that is, at this frequency, the waveguide is practically invisible.

Further optimization of the excitation beam [23] will provide even smaller radiative losses and higher quality factors of the perfect nonradiating modes.

## 2.2 TE polarization

The case of TE polarization is described by the Helmholtz equation for the electric field $E_z$, obtained from (5) by replacing $H_z \to E_z$. Looking for $E_z$ inside and outside the waveguide in a form similar to (6) and (7) and using the boundary conditions for the continuity of the tangential field components [35], we arrive at the following dispersion equation for perfect nonradiating TE modes:

$$J_m\left(k_0R\sqrt{\varepsilon}\right)J'_m\left(k_0R\right) = \sqrt{\varepsilon}J_m\left(k_0R\right)J'_m\left(k_0R\sqrt{\varepsilon}\right). \quad (17)$$

This equation differs from the equation for usual modes by replacing of the Hankel functions with the Bessel functions [35], and its solutions appear as deep minima in the scattered power spectrum.

Figure 6 shows the results of simulation of the scattering of the superposition of TE polarized plane waves on a circular waveguide with $\varepsilon=50$.

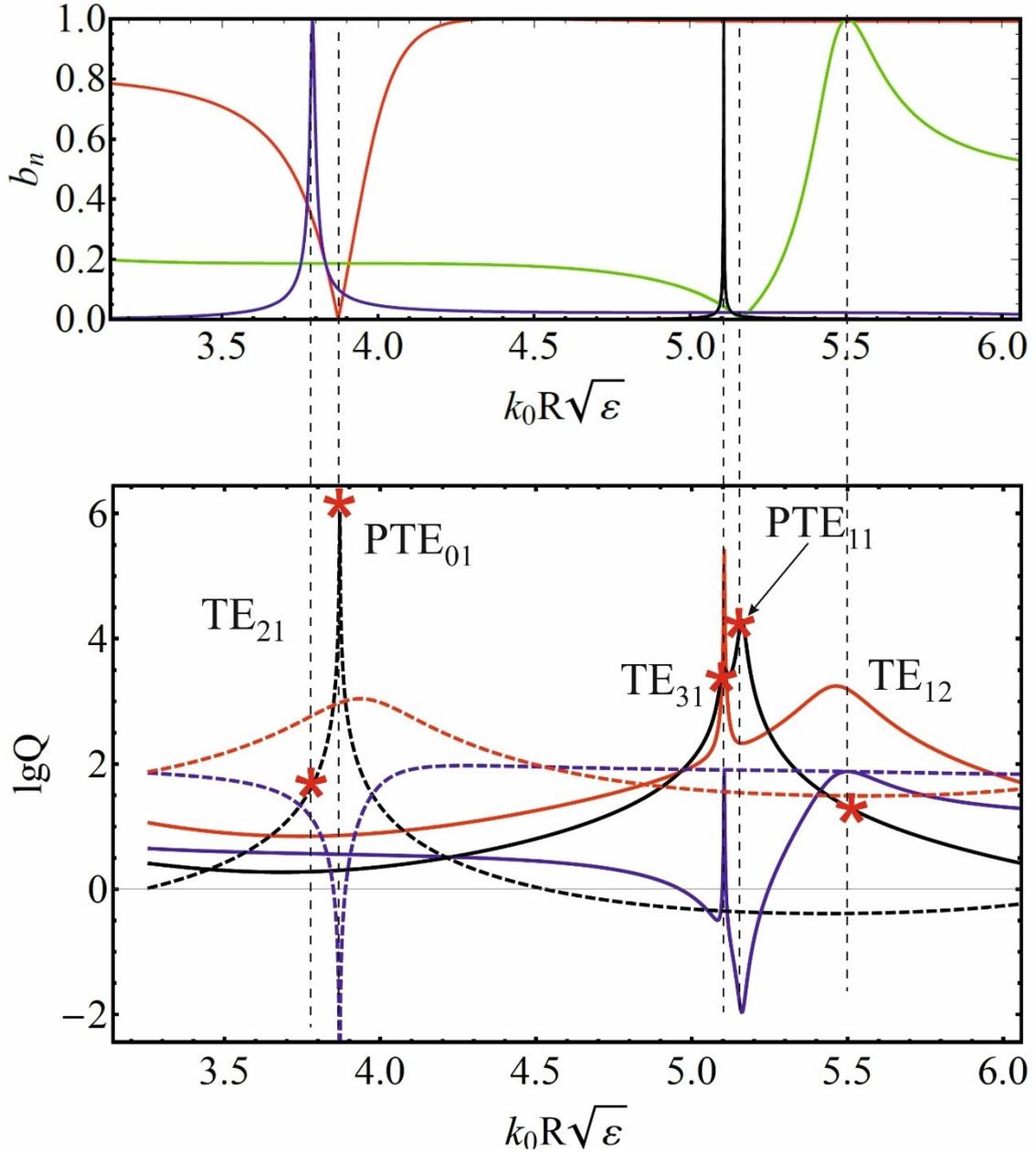

FIG. 6. (Color online) The dependence of (a) Mie coefficient $b_n$ (see [26], p. 199) and (b) scattered power $\lg P^{scat}$ (blue), stored energy $\lg W^{stored}$ (red), and radiative $Q$ factor $\lg Q$ (black) on the size parameter $k_0 R \sqrt{\varepsilon}$ of a waveguide of a circular cross-section obtained within the Comsol Multiphysics software. The solids curves correspond to the excitation field, and the dashed curves correspond to the excitation field $E_z^{exc} = \cos(k_0 x \cos\alpha) \cos(k_0 y \sin\alpha)$, at $\alpha=\pi/4$ and $\varepsilon=50$. The asterisks on the black

curves show the $Q$ factor value of the quasinormal TE$_{21}$, TE$_{12}$, TE$_{31}$ and perfect PTE$_{01}$, PTE$_{11}$ modes. PTE stands for perfect TE nonradiating mode.

Again, FIG.6 shows clearly the appearance of deep minima in the spectrum of the scattered power and extremely high maxima in the spectrum of the radiative $Q$ factor, respectively. These extremums demonstrate distinctly the appearance of high quality perfect PTE$_{01}$ and PTE$_{11}$ modes. Note that $Q$ factor of the PTE$_{01}$ and PTE$_{11}$ mode is substantially greater than the $Q$ factor of usual modes!

The circular cylindrical waveguide is a unique highly symmetrical system where the Helmholtz equation (5) can be exactly solved by the separation of variables. Therefore, a natural question arises about the existence of perfect modes in cylindrical waveguides of other cross-sections. It will be shown in Sections 3 and 4 that in cylindrical waveguides of an arbitrary elliptical cross-section an infinite number of perfect nonradiating modes also exists and they manifest themselves as superdeep minima in the scattered power spectra.

## 3. PERFECT NONRADIATING MODES IN AN ELLIPTICAL WAVEGUIDE: TM POLARIZATION

In this case, it is convenient to solve the Helmholtz equation (5) for TM polarization in the elliptic coordinate system. Cartesian $x$ and $y$ coordinates are related to elliptic coordinates $0 \leq \xi < \infty$, $0 \leq \eta \leq 2\pi$ by the following relations [38]:

$$x = f \cosh\xi \cos\eta, \quad y = f \sinh\xi \sin\eta, \qquad (18)$$

where $f = \sqrt{a^2 - b^2}$ is the half of the focal length of the waveguide ellipse, $a$ and $b$ are the semiaxes of the ellipse (FIG. 1b, $a > b$). The case of a circular waveguide can be obtained in the limit $f \to 0$ and $\xi \to \infty$. In this limit, $f \exp(\xi)/2 \to \rho$ and $\eta \to \varphi$, where $\rho$ and $\varphi$ are the polar coordinates (see section 2.1). The elliptical boundary is described by the equation $\xi = \xi_0 = \operatorname{arccoth}(a/b)$.

The dependence of the eigenfrequencies of perfect nonradiating modes on the shape of the waveguide, that is, on the ratio $t=a/b$, can be compared correctly if the cross-sectional area of circular and elliptical waveguides is constant, that is, when the equality $\pi ab = \pi R^2$ is fulfilled. Therefore, we will use the following expressions for the semiaxes $a$ and $b$, expressed in terms of the shape parameter $t$ and the radius $R$:

$$a = R\sqrt{t}, \quad b = R/\sqrt{t}, \quad t = a/b. \tag{19}$$

An arbitrary nonsingular solution of the homogeneous equation (5) inside ($H_z^{in}$) and outside ($H_z^{out}$) of an elliptical waveguide has the form [38]:

$$H_z^{in} = \sum_{m=0}^{\infty} A_m \mathrm{Ce}_m(\xi,q) \mathrm{ce}_m(\eta,q) + \sum_{m=1}^{\infty} B_m \mathrm{Se}_m(\xi,q) \mathrm{se}_m(\eta,q),$$

$$H_z^{out} = \sum_{m=0}^{\infty} C_m \mathrm{Ce}_m(\xi,q_0) \mathrm{ce}_m(\eta,q_0) + \sum_{m=1}^{\infty} D_m \mathrm{Se}_m(\xi,q_0) \mathrm{se}_m(\eta,q_0), \tag{20}$$

where $\mathrm{ce}_m(\eta,q)$, $\mathrm{se}_m(\eta,q)$ are the elliptic cosines and sines (Mathieu functions), and $\mathrm{Ce}_m(\xi,q)$, $\mathrm{Se}_m(\xi,q)$ are the modified Mathieu functions of the first kind [31,38], $q = (k_0 f)^2 \varepsilon / 4$, $q_0 = q/\varepsilon$ are the size parameters, $A_m$, $B_m$, $C_m$, $D_m$ are the expansion coefficients, that can be found from the boundary conditions at $\xi = \xi_0$.

It is important to note that to determine the dependence of the fields on the elliptic radius $\xi$, both inside and outside the waveguide, we use functions of the first kind $\mathrm{Ce}_m(\xi,q)$, $\mathrm{Se}_m(\xi,q)$, *nonsingular* inside the waveguide and real, that is, they do not lead to an energy flow to infinity. This choice distinguishes fundamentally our solution from the usual solution for radiating quasinormal modes [35,38,39].

Since elliptic cosines and sines have different parities, they describe orthogonal modes. Below, for definiteness, we derive and analyze the dispersion equation for even perfect nonradiating modes, which are described by elliptic cosines $\mathrm{Ce}_m(\xi,q)$.

By equating the tangential components of the electric and magnetic fields inside and outside the waveguide on its boundary ($\xi = \xi_0$) [35,38,39] from (20) we obtain the equations:

$$\sum_{m=0}^{\infty} A_m \mathrm{Ce}_m(\xi_0, q) \mathrm{ce}_m(\eta, q) = \sum_{m=0}^{\infty} C_m \mathrm{Ce}_m(\xi_0, q_0) \mathrm{ce}_m(\eta, q_0),$$
$$\frac{1}{\varepsilon} \sum_{m=0}^{\infty} A_m \mathrm{Ce}'_m(\xi_0, q) \mathrm{ce}_m(\eta, q) = \sum_{m=0}^{\infty} C_m \mathrm{Ce}'_m(\xi_0, q_0) \mathrm{ce}_m(\eta, q_0),$$
(21)

where here and bellow the prime near the Mathieu function means its derivative over the coordinate $\xi$, i.e., $\mathrm{Ce}'_m(\xi_0, q) = \partial \mathrm{Ce}_m(\xi_0, q) / \partial \xi_0$. Multiplying both equations on $\mathrm{ce}_n(\eta, q)$, integrating over $\eta$ from 0 to $2\pi$, and using the orthogonality of elliptic functions instead of (21), we obtain

$$A_n \mathrm{Ce}_n(\xi_0, q) = \sum_{m=0}^{\infty} \Pi c_{nm}(q, q_0) C_m \mathrm{Ce}_m(\xi_0, q_0),$$
$$\frac{1}{\varepsilon} A_n \mathrm{Ce}'_n(\xi_0, q) = \sum_{m=0}^{\infty} \Pi c_{nm}(q, q_0) C_m \mathrm{Ce}'_m(\xi_0, q_0),$$
(22)

where

$$\Pi c_{nm}(q, q_0) = \frac{1}{\pi} \int_0^{2\pi} \mathrm{ce}_n(\eta, q) \mathrm{ce}_m(\eta, q_0) d\eta \qquad (23)$$

stands for the overlap integral of angular elliptical function of different $q$ and $q_0$. FIG.7 shows the dependence of the absolute value of the overlap integral $|\Pi c_{nm}(q, q_0)|$ on the indices $n$, $m$.

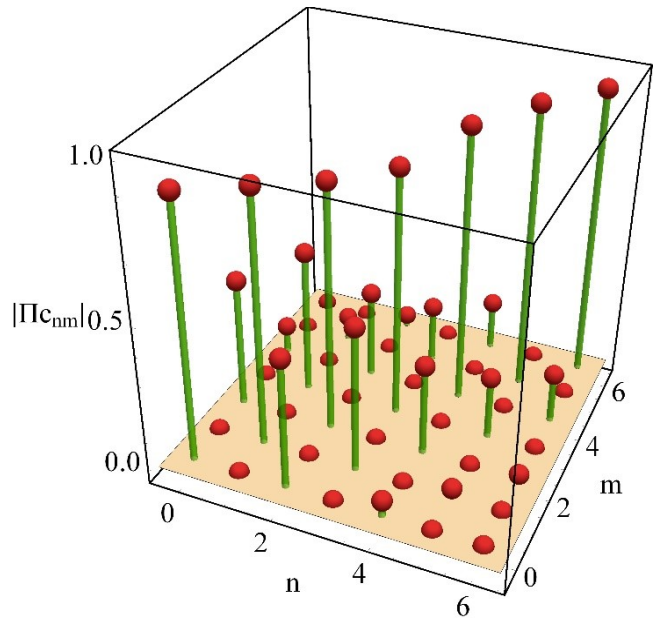

FIG.7. (Color online) The dependence of $|\text{Пc}_{nm}(q,q_0)|$ on $n$, $m$ for $q = 4$, $q_0 = q/\sqrt{\varepsilon}$, $\varepsilon = 50$ is shown by red pinheads.

FIG.7 shows that:

1) only modes with the same parity interact with each other (it follows also from the definition of the Mathieu functions [38,39]);

2) for each mode, interaction is essential only with the nearest neighbors of the same parity $2m \Leftrightarrow 2(m \pm 1)$, $2m+1 \Leftrightarrow 2(m \pm 1)+1$. This circumstance simplifies specific calculations (see below).

Eliminating $A_n$ from (22), we obtain a homogeneous system of equations for $C_m$:

$$\sum_{m=0}^{\infty} \text{Mc}_{nm} C_m = 0, \qquad (24)$$

where

$$\text{Mc}_{nm} = \text{Пc}_{nm}(q,q_0)\left(\text{Ce}_n(\xi_0,q)\text{Ce}'_m(\xi_0,q_0) - \frac{1}{\varepsilon}\text{Ce}'_n(\xi_0,q)\text{Ce}_m(\xi_0,q_0)\right). \qquad (25)$$

The dispersion equation determining the eigenfrequencies of even perfect TM modes takes the form:

$$\det \text{Mc} = 0, \qquad (26)$$

where Mc is the infinite square matrix with matrix elements $\text{Mc}_{nm}$. Note, that since the overlap integral $\text{Пc}_{nm}$ is nonzero only if $n$ and $m$ have the same parity, equation (26) splits into two independent systems of equations: one system for PTM modes with $m,n = 0, 2, 4, 6, \ldots$, and another system for PTM modes with $m,n = 1, 3, 5, \ldots$

For perfect nonradiating modes, which are expanded in elliptic sines, the dispersion equation has a similar form with the replacement of elliptic cosines by elliptic sines

$$\det \text{Ms} = 0, \qquad (27)$$

where the matrix elements of $\text{Ms}_{nm}$ and the overlap integral $\text{Пs}_{nm}$ have the form

$$\mathrm{Ms}_{nm} = \Pi \mathrm{s}_{nm}(q,q_0)\left(\mathrm{Se}_n(\xi_0,q)\mathrm{Se}'_m(\xi_0,q_0) - \frac{1}{\varepsilon}\mathrm{Se}'_n(\xi_0,q)\mathrm{Se}_m(\xi_0,q_0)\right),$$

$$\Pi \mathrm{s}_{nm}(q,q_0) = \frac{1}{\pi}\int_0^{2\pi} \mathrm{se}_n(\eta,q)\mathrm{se}_m(\eta,q_0)d\eta. \qquad (28)$$

The modes that satisfy (26) and (27) will be denoted by the additional letter c or s, that is, as TMc and TMs, respectively.

As in the case of equation (26), instead of (27) there are 2 equations for TMs modes: one for modes with indices 2, 4, 6, …, and another for modes with indices 1, 3, 5, … This is again due to the fact that the overlap integral $\Pi \mathrm{s}_{nm}$ can take nonzero values only if the indices $n$ and $m$ have the same parity.

Our main result is the proof that equations (26), (27) have an infinite number of real roots and therefore, in the case of an elliptical waveguide, there are an infinite number of perfect nonradiating TM modes.

First of all, equations (26) and (27) have an infinite number of real roots at $t=a/b=1$, that is, in the case of a circular cylinder (see Section 2). In this case $f \to 0$, $\Pi \mathrm{c}_{nm}(q,q_0)$, $\Pi \mathrm{s}_{nm}(q,q_0) \to \delta_{nm}$, elliptic functions become cylindrical ones [38,40], and both equations (26) and (27) reduce to equation (8), that has definitely an infinite number of real roots.

Since the eigenvalues depend analytically on the shape parameter $t$ [41], then even for nonzero values of this parameter, equations (25) and (26) will have an infinite number of real solutions. The specific values of eigenfrequencies of perfect nonradiating modes for different waveguide cross-sections were found numerically by the truncation method, that is, we solved a finite system of equations that provides the given accuracy of the solution [39]. FIG. 7 and the results of our calculations show that even 3×3 submatrices of Mc or Ms can be used to calculate eigenfrequencies with sufficiently high accuracy for $1 < a/b < 2$.

FIG. 8 shows *real* solutions of (26) and (27) for arbitrary values of the parameter $t=a/b$.

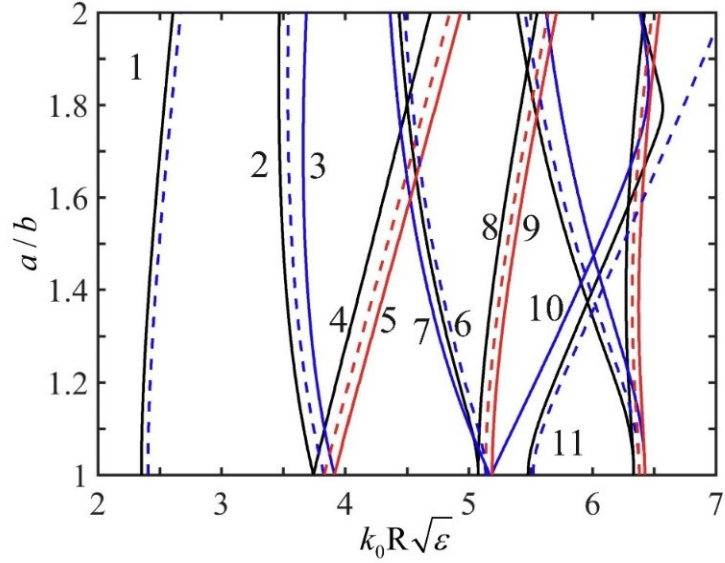

FIG. 8. (Color online) The dependence of eigenfrequencies $k_0 R\sqrt{\varepsilon}$ of the perfect nonradiating TM modes (PTM) on the shape of the ellipse, $a/b$ at $\varepsilon=50$. The solid lines of blue color are the solution of equation (26), the solid lines of red color are the solutions of (27). The dashed blue and green lines are the confined modes in a closed elliptical waveguide [38]. The black color stands for the usual modes. The modes are marked with the numbers: 1- $TMc_{01}$, 2 – $TMc_{11}$, 3 – $PTMc_{11}$, 4 – $TMs_{11}$, 5 – $PTMs_{11}$, 6 – $TMc_{21}$, 7 - hybrid $PTMc_{02,1}$, 8 – $TMs_{21}$, 9 – $PTMs_{21}$, 10– hybrid $PTMc_{20,1}$, 11 – $TMc_{02}$,

Figures 9 and 10 show the dependences of the frequencies of perfect nonradiating modes on the inverse permittivity $1/\varepsilon$.

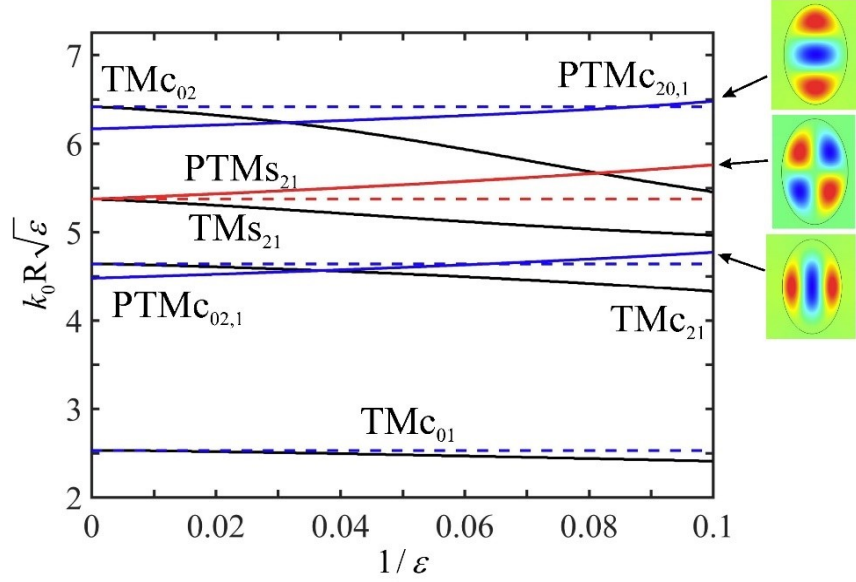

FIG. 9. (Color online) Eigenfrequencies $k_0R\sqrt{\varepsilon}$ of even ((26), blue curves) and odd ((27), red curves) TM perfect nonradiating modes (PTM) and usual modes (black curves) versus $1/\varepsilon$ for the ellipse with $a/b=1.6$, $m=0,2$. The horizontal dashed lines correspond to confined modes, that is, the solution of the equations $\text{Ce}_m(\xi_0,q)=0$ and $\text{Se}_m(\xi_0,q)=0$.

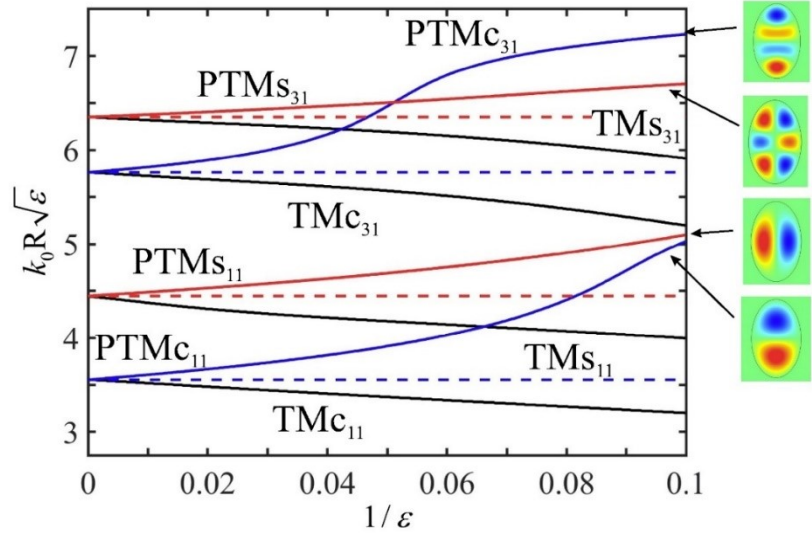

FIG. 10. (Color online) Eigenfrequencies $k_0R\sqrt{\varepsilon}$ of even ((26), blue curves) and odd ((27), red curves) TM perfect nonradiating modes (PTM) and usual modes (black curves) versus $1/\varepsilon$ for the ellipse with $a/b=1.6$ $m=1,3$. The horizontal dashed

lines correspond to the solution of the equations $\text{Ce}_m(\xi_0, q) = 0$ and $\text{Se}_m(\xi_0, q) = 0$.

The dispersion curves in Figs. 8-10 are in complete agreement with the results of Comsol simulation of plane wave scattering by an elliptical waveguide.

It is important to note that not all PTM modes in elliptical waveguides are confined modes in the limit $\varepsilon \to \infty$ [36]. It can be seen from FIG. 9 that even PTMc$_{02,1}$ and PTMc$_{20,1}$ modes are not confined, that is, in the limit $\varepsilon \to \infty$ their frequencies do not tend to the roots of the equation $\text{Ce}_{2m}(\xi_0, q) = 0$, and their fields outside the waveguide do not disappear. In the case of a circular cylinder, only the TM mode with index 0 is not confined (see Section 2). In the case of waveguides with an elliptical cross section, all even PTMc modes interact. It leads to their hybridization and violation of their confinement properties. All usual quasinormal modes are confined. This situation differs significantly from the case of 3D axisymmetric particles, where all perfect nonradiating PTM modes are confined [29,30].

The found perfect nonradiating TM modes in elliptical cylinders are not abstract solutions of sourceless Maxwell's equations. They are of great practical importance for finding the conditions for extremely small or even zero scattered power at a finite stored energy, leading to the unlimited radiative $Q$ factor.

Indeed, when considering the scattering problem, any incident field can be expanded over elliptic functions [38,39]. Particularly in this case, the even excitation and scattered fields outside the waveguide can be represented as

$$H_z^{exc} = \sum_{m=0}^{\infty} C_m^{exc} \text{Ce}_m(\xi, q) \text{ce}_m(\eta, q), \tag{29}$$

$$H_z^{scat} = \sum_{m=0}^{\infty} C_m^{scat} \text{Me}_m^{(1)}(\xi, q_0) \text{ce}_m(\eta, q_0), \tag{30}$$

where $\text{Me}_m^{(1)}(\xi, q_0)$ is the elliptic function similar to the Hankel function of the first kind [38,39], and $C_m^{exc}$, $C_m^{scat}$ are the expansion coefficients found from the boundary conditions.

Repeating the procedure described above for finding perfect nonradiating

modes one can see that the coefficients $C_m^{scat}$ for scattered fields can be found using the relationship

$$C_p^{scat} = -\sum_{m=0}^{\infty} \left(\bar{M}c\right)^{-1}_{pm} \sum_{n=0}^{\infty} Mc_{mn} C_n^{exc}, \qquad (31)$$

where

$$\bar{M}c_{nm} = \Pi c_{nm}(q,q_0) \left[ Me_n^{(1)}(\xi_0,q_0) Ce'_m(\xi_0,q_0) - \frac{1}{\varepsilon} Me_n^{(1)\prime}(\xi_0,q_0) Ce_m(\xi_0,q_0) \right], \qquad (32)$$

and the expression for $Mc_{nm}$ is given in (25).

If the coefficients $C_m^{exc}$ are a nontrivial solution of (24), that is, they are eigenvectors of perfect nonradiating modes, then the scattered field and the scattered power will be exactly zero at the frequencies of perfect nonradiating modes! It means that the elliptical cylinder will be invisible at the frequencies of the perfect nonradiating modes. In the case of excitation fields of other parity, the situation is completely analogous.

In practice, it is hardly possible to create an excitation field with a structure exactly equal to the structure of the external field of a perfect nonradiating mode, and therefore with the help of Comsol Multiphysics software we have considered how effectively perfect nonradiating modes manifest themselves when plane waves are scattered by elliptical waveguides of different cross-sections.

FIG. 11. shows the results of numerical calculations of scattering spectra when TM polarized plane waves excite an elliptical cylinder.

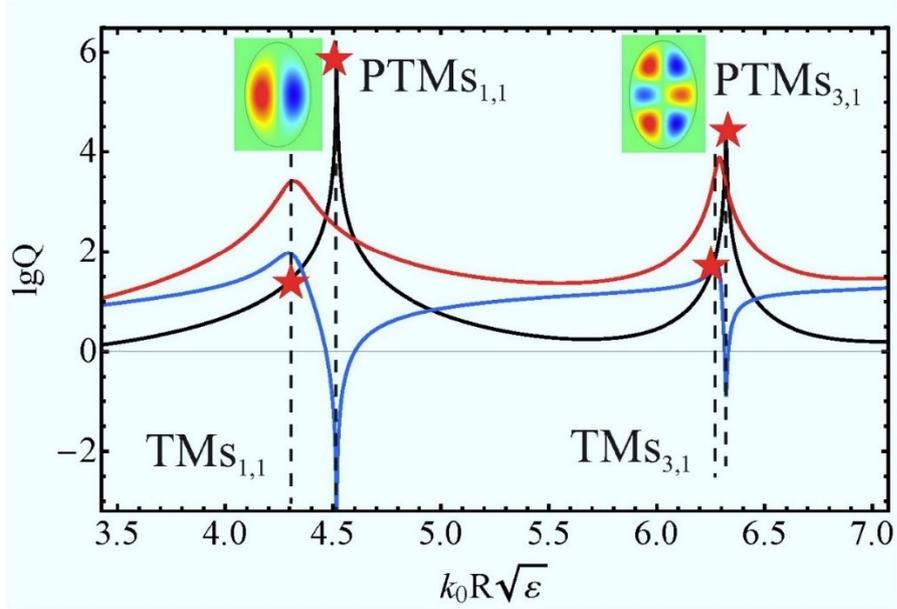

FIG. 11. (Color online) The dependence of scattered power $\lg P^{scat}$ (blue), stored energy $\lg W^{stored}$ (red), and radiative $Q$ factor $\lg Q$ (black) on the size parameter $k_0 R\sqrt{\varepsilon}$ of a waveguide with $a/b$=1.6 obtained within the Comsol simulation. The solid curves correspond to the excitation field $H_z^{exc} = \sin(k_0 x \cos\alpha)\cos(k_0 y \sin\alpha)$, $\alpha=\pi/4$, $\varepsilon$=50. The asterisk on the black curve shows the $Q$ factor value of the PTMs$_{11}$, PTMs$_{31}$, TMs$_{11}$, TMs$_{31}$ modes. PTM stands for perfect TM nonradiating mode.

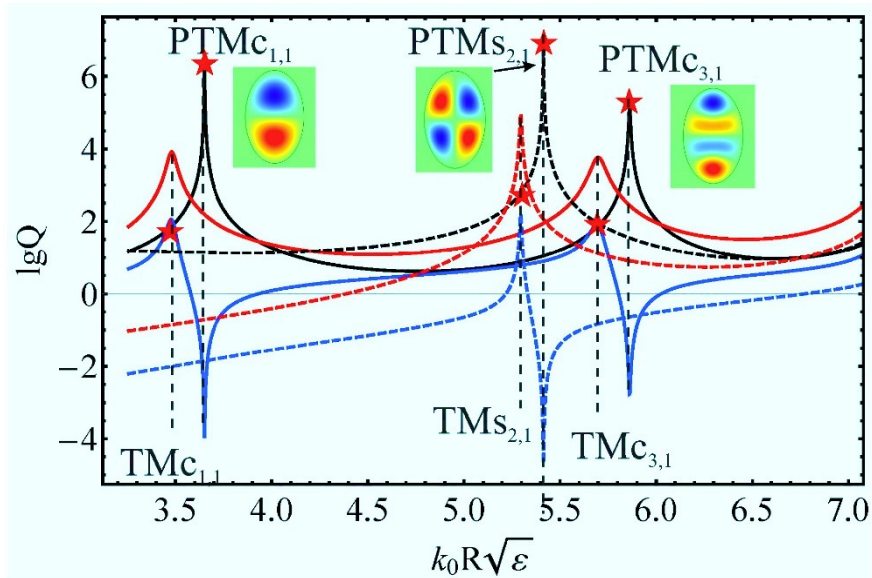

FIG. 12. (Color online) The dependence of scattered power $\lg P^{scat}$ (blue), stored energy $\lg W^{stored}$ (red), and radiative $Q$ factor $\lg Q$ (black) on the size parameter $k_0 R\sqrt{\varepsilon}$ of a waveguide with $a/b=1.6$ obtained within the Comsol simulation. The solid curves correspond to the excitation field $H_z^{exc} = \cos(k_0 x \cos\alpha)\sin(k_0 y \sin\alpha)$, the dashed curves correspond to the excitation field $H_z^{exc} = \sin(k_0 x \cos\alpha)\sin(k_0 y \sin\alpha)$, $\alpha=\pi/4$, $\varepsilon=50$. The asterisk on the black curve shows the $Q$ factor value of the modes.

FIG.11, 12 show clearly the appearance of extremely high quality perfect PTMs$_{11}$, PTMs$_{31}$, PTMc$_{11}$, and PTMc$_{31}$ modes in the spectra of a power scattered by an elliptical waveguide. Note that $Q$ factor of the PTMc$_{11}$, PTMs$_{11}$ modes is by six orders of magnitude greater than $Q$ factor of a usual quasinormal TMc$_{11}$, TMs$_{11}$ modes! The minima of the blue lines (and the maxima of the black ones) correspond to the perfect modes. They are shown by the dashed lines in FIG. 11, 12.

Although excitation of an elliptical cylinder by a plane standing wave gives deep minima in the scattered power at frequencies corresponding to the frequencies of perfect nonradiating modes, further optimization of the excitation fields makes it possible to reduce the scattered power further. Such an optimization can be performed, in particular, using the method described in [23], implying that the excitation field can be considered as a superposition of plane waves, that, in turn, can be expanded in terms of elliptic functions [38,39]:

$$H_z^{exc} = \sum_j \beta_j \sin(k_0 x \cos\alpha_j)\cos(k_0 y \sin\alpha_j)$$
$$= 2\sum_{m=0}^{\infty} \frac{\overline{C}_m}{p_{2m+1}} \mathrm{Ce}_{2m+1}(\xi,q_0)\mathrm{ce}_{2m+1}(\eta,q_0),$$
(33)

where

$$\overline{C}_m = \sum_j \beta_j \mathrm{ce}_{2m+1}(\alpha_j, q_0).$$
(34)

By tuning the parameters $\alpha_j$ and $\beta_j$, it is possible to bring the excitation field closer to the external field of the perfect nonradiating mode (20) and thereby ensure arbitrarily deep minima of the scattered power and arbitrarily high values of the radiative $Q$ factor.

## 4. PERFECT NONRADIATING MODES IN A CYLINDRICAL ELLIPTICAL WAVEGUIDE: TE POLARIZATION

In this case, the solution of the Helmholtz equation for the component $E_z$ inside ($E_z^{in}$) and outside ($E_z^{out}$) of the waveguide can be sought in the form:

$$E_z^{in} = \sum_{m=0}^{\infty} A_m \mathrm{Ce}_m(\xi,q)\mathrm{ce}_m(\eta,q) + \sum_{m=1}^{\infty} B_m \mathrm{Se}_m(\xi,q)\mathrm{se}_m(\eta,q),$$
$$E_z^{out} = \sum_{m=0}^{\infty} C_m \mathrm{Ce}_m(\xi,q_0)\mathrm{ce}_m(\eta,q_0) + \sum_{m=1}^{\infty} D_m \mathrm{Se}_m(\xi,q_0)\mathrm{se}_m(\eta,q_0),$$
(35)

where $A_m$, $B_m$, $C_m$, $D_m$ are the expansion coefficients that can be found from the boundary conditions at $\xi = \xi_0$.

Repeating for (35) the calculations given in the previous section, we can obtain the following equations for the TE modes:

$$\sum_{m=0}^{\infty} \mathrm{Nc}_{nm} C_m = 0, \tag{36}$$

$$\sum_{m=1}^{\infty} \mathrm{Ns}_{nm} D_m = 0. \tag{37}$$

where

$$\mathrm{Nc}_{nm} = \Pi c_{nm}(q,q_0)\left[\mathrm{Ce}_n(\xi_0,q)\mathrm{Ce}'_m(\xi_0,q_0) - \mathrm{Ce}'_n(\xi_0,q)\mathrm{Ce}_m(\xi_0,q_0)\right], \tag{38}$$

$$\mathrm{Ns}_{nm} = \Pi s_{nm}(q,q_0)\left[\mathrm{Se}_n(\xi_0,q)\mathrm{Se}'_m(\xi_0,q_0) - \mathrm{Se}'_n(\xi_0,q)\mathrm{Se}_m(\xi_0,q_0)\right]. \tag{39}$$

Equations (36) and (37) have nontrivial solutions if

$$\det \mathrm{Nc} = 0 \tag{40}$$

$$\det \mathrm{Ns} = 0, \tag{41}$$

Figure 13 shows the dependencies of the eigenfrequencies $k_0 R\sqrt{\varepsilon}$ of perfect nonradiating TE modes, where (40) and (41) have nontrivial solutions, as a function of $t=a/b$, determining the shape of the elliptical cross-section of the waveguide.

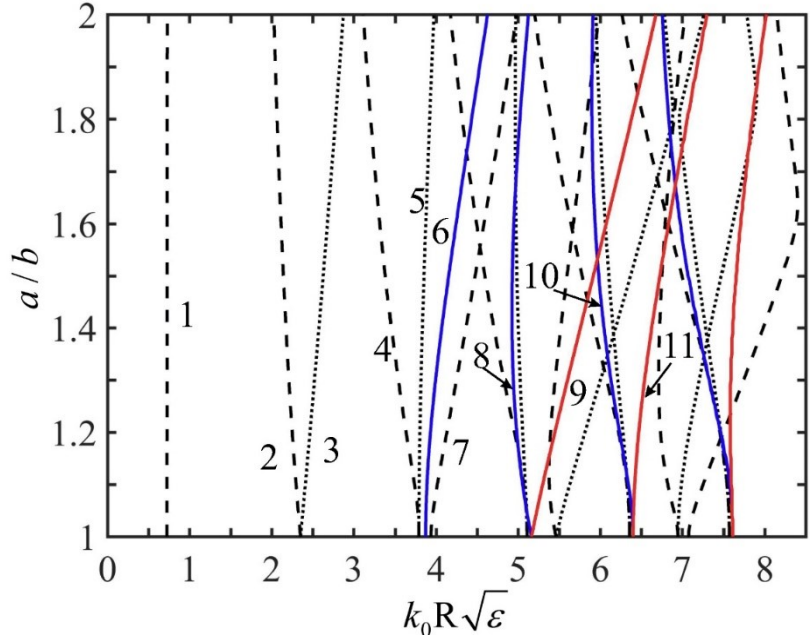

FIG. 13. (Color online) The eigenfrequencies $k_0R\sqrt{\varepsilon}$ of perfect nonradiating TE modes (PTE) at $\varepsilon=50$ as the functions of the aspect ratio $a/b$. The solid blue curves are solutions of (40), the red curves are solutions of (41). The dashed and dotted black curves are the positions of usual TE modes. The curves are marked with the numbers: 1- $TEc_{01}$, 2 – $TEc_{11}$, 3 – $TEs_{11}$, 4 – $TEc_{21}$, 5 – $TEs_{21}$, 6 – $PTEc_{01}$, 7 - $TEc_{02}$, 8 – $PTEc_{11}$, 9 – $PTEs_{11}$, 10-$PTEc_{21}$, 11 – $PTEs_{21}$,

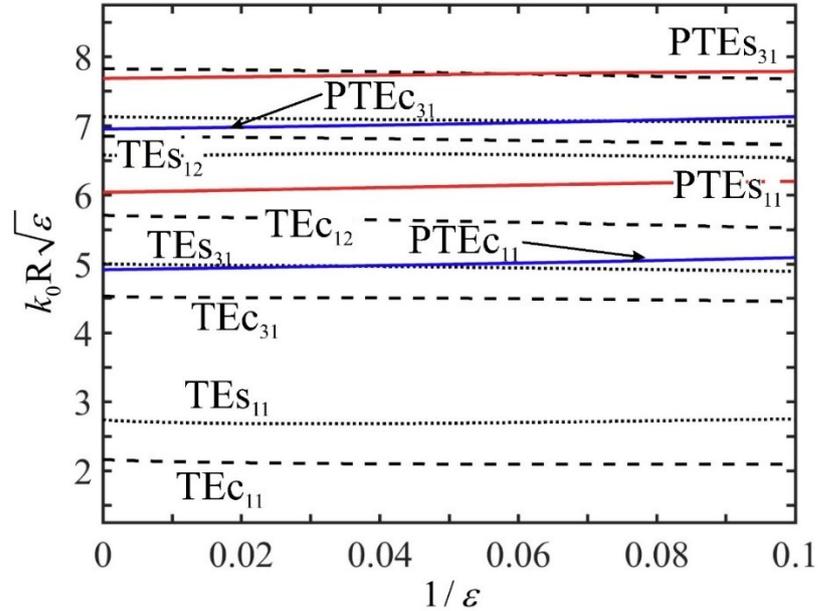

FIG. 14. (Color online) Dependence of the frequencies $k_0R\sqrt{\varepsilon}$ of odd perfect TE

modes (PTE) on $1/\varepsilon$ for the ellipse with $a/b=1.6$. The blue curves correspond to the solutions of (40), and the red curves correspond to the solution of (41). The dashed and dotted black lines are corresponding to the usual TE modes.

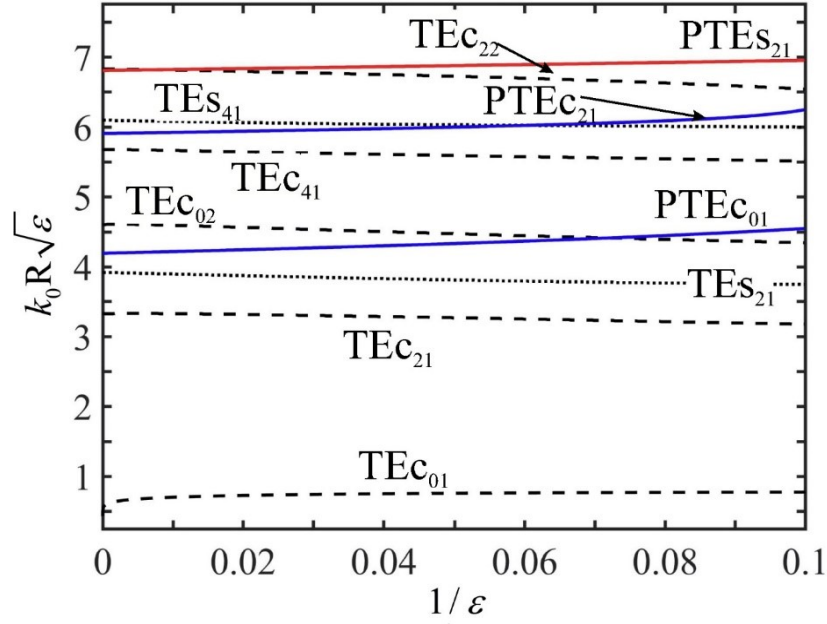

FIG. 15. (Color online) The dependence of the frequencies $k_0 R\sqrt{\varepsilon}$ of even perfect TE modes (PTE) on $1/\varepsilon$ for the ellipse with $a/b=1.6$. The blue curves correspond to the solutions of (40), and the red curves correspond to the solution of (41). The dashed and dotted black lines correspond to the usual modes.

From the analysis of FIGS. 13-15, we can conclude that in the case of TE polarization there is an infinite number of perfect nonradiating modes in an elliptical waveguide.

Experimental observation of perfect nonradiating TE modes in nontrivial elliptical waveguides is more difficult than in the case of TM modes, because:

1) TE modes are not confined [36] even in the limit $\varepsilon \to \infty$,

2) perfect TE modes of orders $m$ and $m+2$ have very close frequencies for any $m=0, 1, 2...$

Nevertheless, using the optimal superposition of plane waves described in Section 3, it is possible, in principle, to excite TE perfect nonradiating modes efficiently.

# 5. APPLICATION OF PERFECT NONRADIATING MODES TO DESIGN DISPLACEMENT SENSORS

Extremely high quality factor and extremely low scattered power at the frequencies of perfect nonradiating modes can be naturally used to develop sensors for various purposes.

For example, if a nanowaveguide of a circular cross section with the center at $x=0$, $y=0$ is excited by the field of a standing wave

$$H_z^{exc} = \sin(k_0 x \cos\alpha)\cos(k_0 y \sin\alpha), \qquad (42),$$

then the scattered power will be determined by the expression [37]:

$$P_0^{scat} = \frac{c}{2\pi k_0} \sum_{n=-\infty}^{\infty} |a_n|^2 \cos^2(n\alpha)\sin^2\left(\frac{n\pi}{2}\right), \qquad (43)$$

where $a_n$ are the Mie scattering coefficients (15). At the frequency of the perfect nonradiating mode $PTM_{11}$, a deep minimum can be observed in the scattered power spectrum (see FIG. 4) and the nanofiber is almost invisible. If, at the frequency of the perfect mode, the cylinder is shifted along the $x$ axis on a small distance $\Delta x$ (see FIG. 16), then the scattered power will be determined by the expression:

$$P_\Delta^{scat} = \frac{c}{2\pi k_0} \sum_{n=-\infty}^{\infty} |a_n|^2 \cos^2(n\alpha)\left[\sin^2\left(\frac{n\pi}{2}\right) + (k_0 \Delta x \cos\alpha)^2 \cos(n\pi)\right]$$
$$= P_0^{scat} + (k_0 \Delta x \cos\alpha)^2 P_1^{scat}. \qquad (44)$$

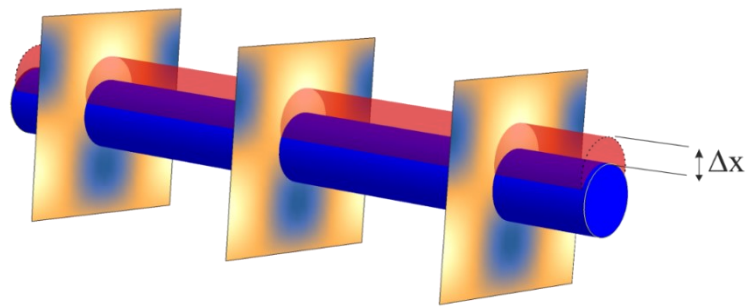

FIG. 16. (Color online) Scheme of the displacement sensor at the frequency of the perfect nonradiating mode. The field distribution of the excitation field (42) is shown in vertical slices (see also FIG. 5).

With such a displacement, the scattered power will double, i.e.

$$P_\Delta^{scat} = 2P_0^{scat}, \qquad (45)$$

if $(\alpha = \pi/4,\ \varepsilon = 50)$

$$\frac{\Delta x}{R} = \frac{1}{k_0 R \cos\alpha}\sqrt{\frac{P_0^{scat}}{P_1^{scat}}} \approx 0.001, \qquad (46)$$

that is, with the help of such modes, it is possible to detect the displacement of a cylinder with a radius of 50 nm by a value of the order of 50 nm/1000=0.5 Å.

Similarly, perfect nonradiating modes can be used to detect a change in the effective radius of a fiber functionalized with an antibody layer when recognizable proteins are bound to it.

## 6. CONCLUSION

In conclusion, we have developed the concept of perfect nonradiating eigenmodes of light in dielectric nanofibers. These modes are exact solutions of the sourceless Maxwell equations, and the number of these modes is unlimited.

The physics of perfect nonradiating modes is fully different from quasinormal modes and has no analogues. The perfect modes are closest to strange Neumann-Wigner modes [42], but unlike the latter, the optical potential of a nanoparticle (permittivity) differs from vacuum value only in a bounded region of space, distinguishing perfect nonradiating modes from Neumann-Wigner modes [42] fundamentally.

Due to extremely small scattered power ("invisibility") and unlimited radiative $Q$ factors, our finding paves the way for development of new nano-optical devices with high concentration of field inside nanoparticles and extremely small radiative losses, including low threshold nanolasers, biosensors, displacement sensors, parametric amplifiers, and nanophotonics quantum circuits

**Declaration of Competing Interest**

The authors report no declarations of interest.

**Acknowledgement**

Funding by the Russian Foundation for the Basic Research (grant 20-12-50136) is acknowledged by V.K.